\documentclass[final,5p,times,twocolumn,numbers]{elsarticle}

\usepackage{amssymb}
\usepackage{lipsum}

\journal{Nuclear Physics A}

\usepackage{hyperref}
\hypersetup{
    colorlinks=true,
    linkcolor=blue,
    filecolor=blue,
    urlcolor=blue,
    citecolor=blue
}

\begin{document}
\begin{frontmatter}

\title{Polarizing $^3$He via Metastability Exchange Optical Pumping Using a 1.2 mbar Sealed Cell at Magnetic Fields up to 5 T}

\author[a]{P.~Pandey \corref{cor1}}
\affiliation[a]{organization={Laboratory for Nuclear Science, Massachusetts Institute of Technology}, 
                city={Cambridge}, 
                postcode={02139}, 
                state={MA}, 
                country={USA}}

\author[c]{H.~Lu \corref{cor1}}
\affiliation[c]{organization={University of Tennessee}, 
                city={Knoxville}, 
                postcode={37996}, 
                state={TN}, 
                country={USA}}
 \author[b]{J.~D.~Maxwell}
\affiliation[b]{organization={Thomas Jefferson National Accelerator Facility}, 
                city={Newport News}, 
                postcode={23606}, 
                state={VA}, 
                country={USA}}               
\author[b]{J.~Brock}
\author[b]{C.~D.~Keith}

\author[d]{X.~Li}
\affiliation[d]{organization={Shandong University}, 
                state={Shandong}, 
                postcode={250100}, 
                country={China}}

\author[a]{R.~G.~Milner}
\author[c]{D.~Nguyen}

\cortext[cor1]{Corresponding authors: ppandey@mit.edu (P. Pandey), hlu17@utk.edu (H. Lu).}

\begin{abstract}
We report high nuclear polarization of 1.2 \,mbar $^3$He gas in a sealed cell in magnetic fields up to 5\,T using Metastability Exchange Optical Pumping (MEOP). The creation of a highly polarized $^3$He gas target for use in the 5\,T field of Jefferson Lab's CLAS12 spectrometer would enable new studies of spin-dependent asymmetries on the neutron. A systematic study was conducted to evaluate the effects of discharge intensity, pump laser power, and optical pumping transition schemes on nuclear polarization and pumping rates. Steady-state polarizations up to 86\% in magnetic fields between 2 and 5 T were achieved, with a discharge-on relaxation time of 898 s at 5 T. These results underscore the potential of MEOP for high-field applications in nuclear physics experiments.
\end{abstract}


\end{frontmatter}

\section{Introduction}
\label{introduction}
Polarized $^3$He gas has become a valuable tool as target nuclei for scattering experiments, providing crucial insights into the spin-dependent structure functions of the neutron. In $^3$He, the two protons predominantly align their spins oppositely, effectively negating their influence on the nucleus’s total spin. This leaves the unpaired neutron as the dominant contributor to the nuclear spin, allowing polarized $^3$He to serve as a proxy for a ``polarized neutron target." 

Metastability Exchange Optical Pumping (MEOP) facilitates the hyperpolarization of $^3$He gas nuclei within a homogeneous magnetic field through the interaction of circularly polarized light with an RF-induced plasma discharge \cite{Colegrove_Schearer_Walters_1963}. The RF discharge elevates a small population of $^3$He atoms to the metastable $2^3S_1$ state, which serves as an intermediary for optical pumping. Using 1083 nm laser light, transitions are induced between the $2^3S_1$ and $2^3P_0$ states, selectively altering the magnetic quantum number ($\Delta$m = $\pm$1) depending on the helicity of the incident light. The optical polarization imparted to the metastable atoms is subsequently transferred to the ground-state population via metastability exchange collisions, culminating in macroscopic nuclear polarization of the $^3$He gas.

Over the past 30 years, optically-pumped $^3$He gas targets have been seen extensive use in spin-structure studies at Jefferson Lab using Spin-Exchange polarization techniques (SEOP) \cite{gentile_optically_2017}.
These traditional polarized $^3$He targets are known to function efficiently at high pressures (up to 13 atm) in low magnetic fields (around 10$^{-3}$ T); however, they have not been used to produce polarization in high magnetic field environments. Pioneering advancements in high-field metastability exchange optical pumping begun at the Kastler-Brossel Laboratory (LKB) in Paris \cite{courtade_magnetic_2002} have inspired new applications of polarized $^3$He gas for particle physics experiments requiring strong magnetic fields, such as a polarized ion source for the Electron-Ion Collider (EIC) \cite{zelenski_optically_2023} and a target for the CLAS12 spectrometer at Jefferson Lab \cite{maxwell_concept_2021}. 

A double-cell polarized $^3$He target, designed and constructed in the late 1980's by Caltech for use at MIT-Bates \cite{milner_polarized_1989}, utilized MEOP at low field to produce a polarized target for electron scattering. In this system, the optical pumping cell was held at room temperature, and the electron beam interacted with gas in a separate target cell, which was cooled to cryogenic temperatures to increase its density.  The two cells communicated via diffusion through a short, interconnecting tube. At Jefferson Lab, we are developing a new polarized $^3$He target for the CLAS12 spectrometer \cite{burkert_clas12_2020} following the design of the MIT-Bates target, while incorporating the recent advancements in high-field MEOP. This approach seeks to produce polarized $^3$He within the 5 T solenoid of CLAS12, while maintaining a target thickness similar to that of traditional low-field polarized $^3$He targets. This would enable the CLAS12 detector to perform comprehensive spin-dependent electron scattering measurements across the entire kinematic range, including elastic, quasielastic, resonance, deep inelastic, and deeply virtual exclusive scattering. 

We are actively developing this new CLAS12 target, exploring polarization techniques in experimental conditions. In this paper, we discuss $^3$He polarization achieved at 5\,T in a sealed cell at a pressure of 1.2 mbar. The dimensions of the sealed cell, 2\,cm in diameter and 10\,cm in length, have been designed to match those of the pumping cell intended for ultimate use under the experimental conditions in CLAS12.

\section{Apparatus setup}

The apparatus is centered around two laser systems: \textbf{the optical pumping system} and \textbf{ the optical polarimeter}.  The pumping laser is an Azurlight diode laser, tunable from 1082.6 to 1083.8 nm, with a spectral broadening stage to produce a clean, 2\,GHz wide wavelength peak to match the $^3$He absorption lines. The optical pumping system optics circularly polarize and expand the laser to cover the full cross-section of the cell. 

The probe laser is a Toptica DFB pro L at 1083\,nm that produces up to 70\,mW. The optical polarimeter's probe laser passes through the cell to a mirror, where it is returned back through the cell to a photodiode. To minimize local saturation of the transitions probed, the probe laser intensity must be limited, so the beam is expanded and attenuated to below 1\,mW, as low as achievable with good signal to noise.

\begin{figure}[h]
\centering
\includegraphics[width=\columnwidth]{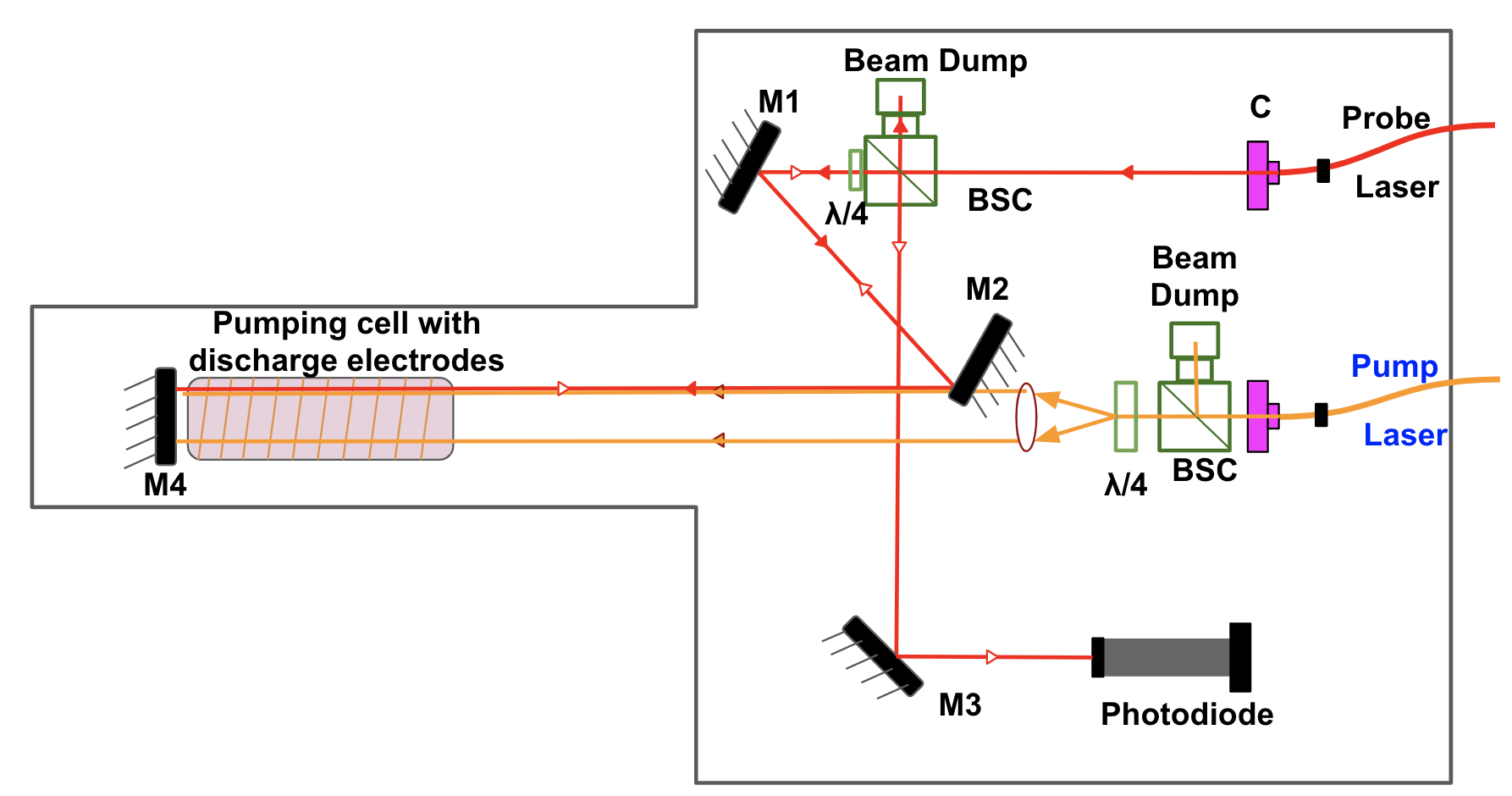}
\caption{The figure shows the optical setup for polarization tests at high field. On the probe laser path, three mirrors are incorporated. The probe laser first passes through a beam-splitter (BSC), then through a quarter-wave plate ($\lambda$/4), and strikes mirror M1. M1 reflects the beam towards M2, and from M2, it is directed to M4. The beam reflected from M4 returns to M2, which then reflects it to M1. The beam then travels to the BSC, reflects off it to M3, and finally reaches the photodiode.}
\label{fig:setup}
\end{figure}

As seen in Fig.~\ref{fig:setup}, both lasers are incident on a sealed, cylindrical borosilicate glass cell filled to 1.2\,mbar with $^3$He. The cell is wrapped with copper wire, which acts as an electrode to produce a discharge plasma in the $^3$He gas. A Thorlabs laser enclosure surrounds the laser optics and a modified cylindrical extension to the enclosure holds the cell within the center of a warm-bore superconducting solenoid magnet.

For further details on the probe acquisition and analysis, please refer to earlier work with this apparatus \cite{li_metastability_2023}.

The superconducting magnet generates a homogeneous magnetic field of up to 5 T within the central region of its 76-cm-long and 13-cm-diameter warm bore. For this work, the magnet is operated at 2, 3, and 5 T to conduct high-field MEOP tests on polarized $^3$He.

\begin{figure}[h!]
    \centering
    \includegraphics[width=0.8\columnwidth]{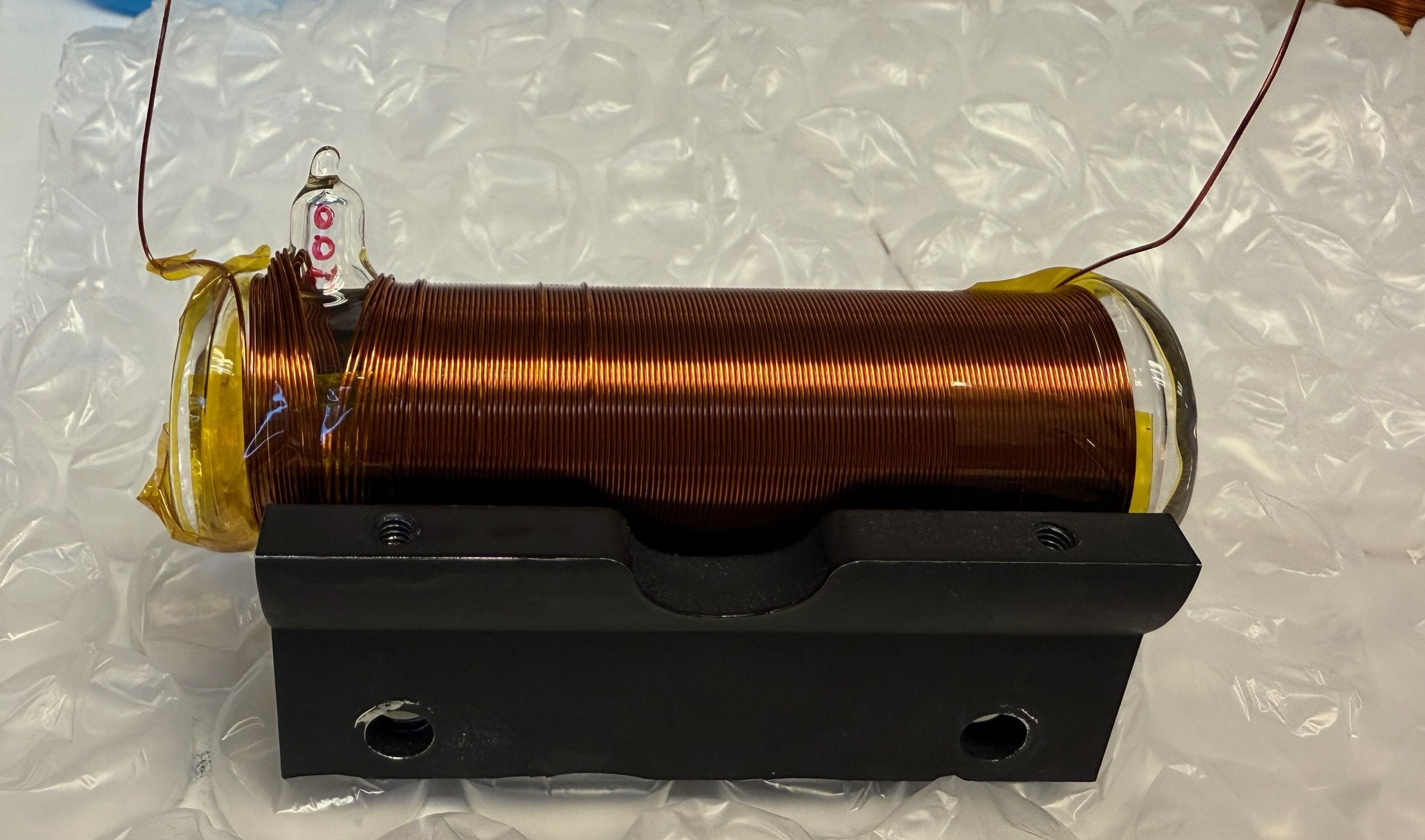}
    \hfill
    \includegraphics[width=0.8\columnwidth]{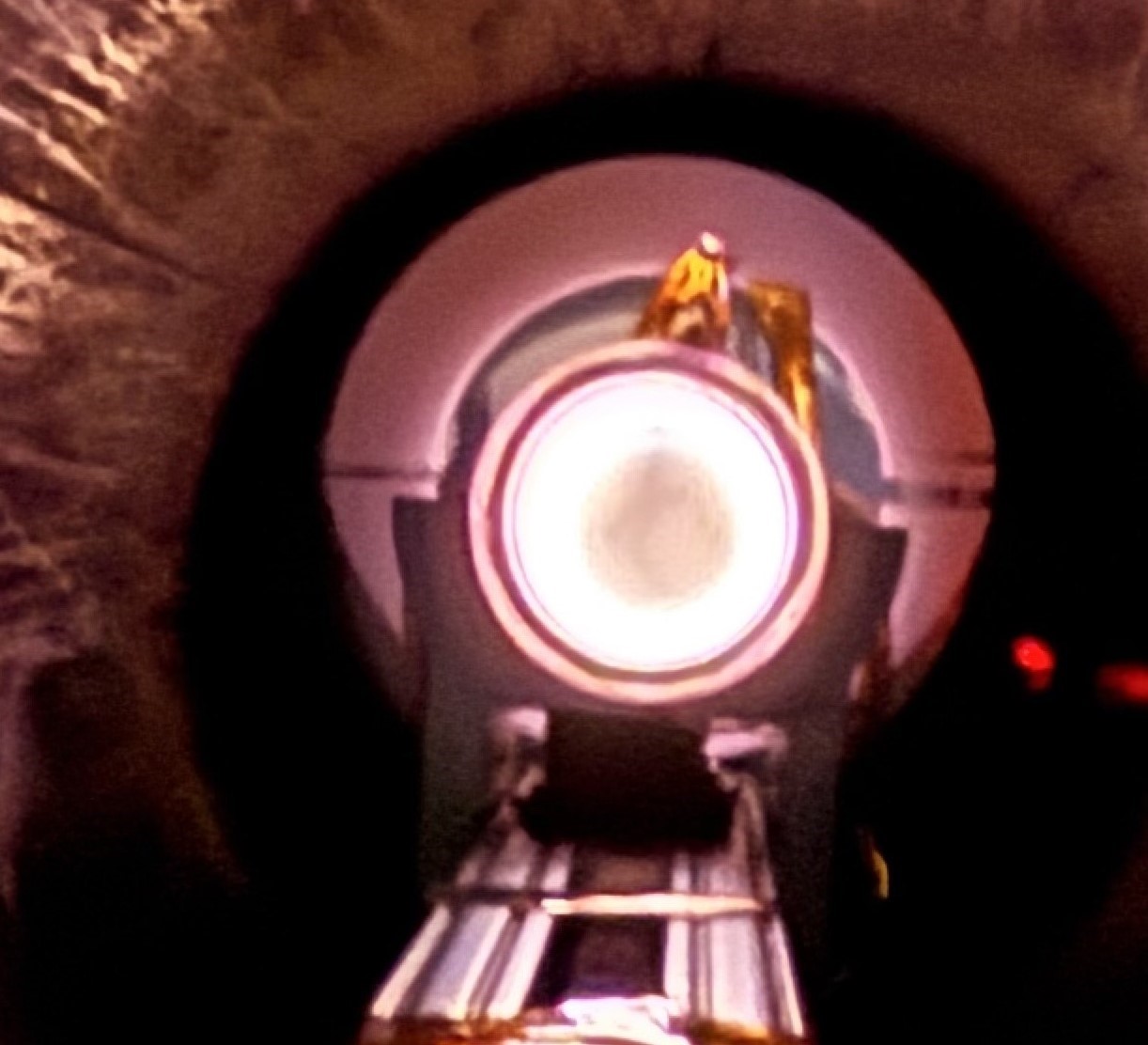}
    \caption{Top: ``Bone'' style cell with 30\,mm windows and a 25.4\,mm diameter volume, showing the inductive electrode wound the length. Bottom: Discharge in the cell gets pushed to the edge while in the 5 T magnetic field.}
    \label{fig:celldischarge}
\end{figure}
A cylindrical, bone-shaped borosilicate glass cell, measuring 10 cm in length, 30 mm windows, and 25.4 mm in diameter, is filled with pure $^3$He gas sealed at a pressure of 1.2 mbar. Before filling and sealing, the cell underwent a cleaning process at Jefferson Lab following Collier’s procedure \cite{collier_metastability_2011}. This involved heating the cell in an ultrasonic bath containing water mixed with detergent for 15 minutes, followed by two rinses with hot water, each lasting 8 minutes. The cell was then sequentially rinsed with acetone, isopropanol, and ethanol before being left to air dry naturally. 

The cell was baked at 200$^\circ$C to remove contaminants, followed by a series of pump-and-purge cycles with $^4$He to eliminate residual gases. Subsequently, $^3$He was introduced and cycled through the cell until the $^3$He discharge spectrum---as measured with an optical spectrometer---appeared clean and free of impurities. Continuous monitoring was performed using a residual gas analyzer (RGA), while the cleanliness of the $^3$He spectrum was consistently assessed throughout the conditioning process. This procedure took approximately one month. One day prior to sealing, the cell was filled with 1.2 mbar of $^3$He and left open to a SAES St-172 based sintered porous getter overnight to further enhance purity. Then the cell was valved off, disconnected from the system, and shipped to Finkenbeiner Inc. in Waltham, MA, for sealing before returning to Jefferson Lab.

The electrical plasma discharge within the cell was generated using electrodes spirally wrapped around the outer surface of the cell, driven by a Stanford Research Systems (SRS) generator and amplified by a 25\,W RF amplifier. We tested several electrode schemes, including capacitive strips and loosely wound inductors, and in our configuration the discharge has been easiest to control with a tightly wound, inductive electrode, as seen in Fig.~\ref{fig:celldischarge}. The discharge amplitude was modulated by a signal generator, providing a reference for the lock-in amplifier which measured the probe laser photodiode amplitude, as in reference \cite{maxwell_enhanced_2020}. The output of the lock-in amplifier was recorded and processed using a Python-based software interface.

The primary update in this setup compared to our previous work \cite{li_metastability_2023} is the more sophisticated probe laser path. Inside the magnetic field, the discharge is pushed to the edges of the cell, see Fig.~\ref{fig:celldischarge}. To ensure the probe laser passes through the portion of the cell with the high concentration of plasma, we incorporated three additional mirrors and a circular polarizer to send the beam in and out of the cell on the same path. The returning beam is then split off to be sent back to the photodiode. This adjustment allowed for precise alignment of the beam profile with the discharge distribution, as depicted in Fig.~\ref{fig:setup}. 

\subsection{Pumping and Polarimetry}
\begin{figure}[]
    \centering
    \includegraphics[width=\columnwidth]{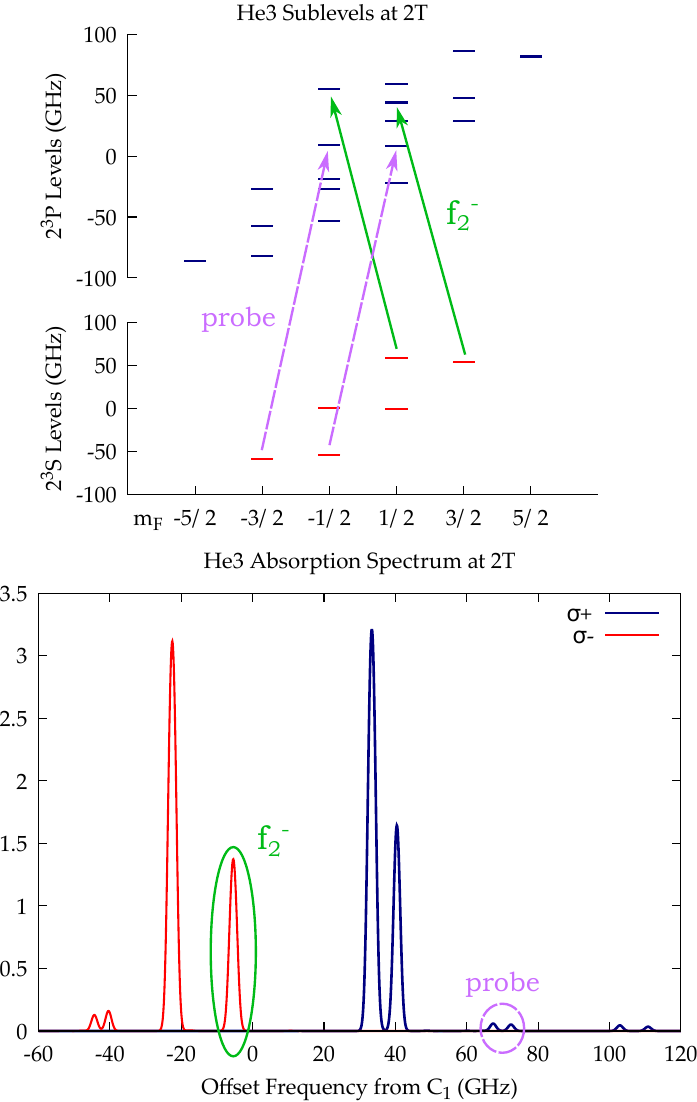}
    \caption{Calculated 2S and 2P sublevels for $^3$He, and absorption peaks at 2\,T from code by P.-J.~Nacher at LKB. The absorption spectrum shows the 4 large pump peaks, from left to right f$_4^-$, f$_2^-$, f$_4^+$, and f$_2^+$. Here, f$_2^-$ is highlighted along with the $\sigma+$ probe peaks used to measure the polarization when pumping that line. The sublevels plot highlights with arrows the sublevel transitions corresponding to those f$_2^-$ and $\sigma+$ probe peaks.}
    \label{fig:levels}
\end{figure}

Both pump and probe laser systems are wavelength tunable to address the major absorption lines of $^3$He at their full separation at 5\,T. There are four main pumping peaks available, two for each handedness of circular polarization and named for the number of unresolved transitions in the line: f$_4^-$, f$_2^-$, f$_4^+$, and f$_2^+$. Each of these lines addresses either 2 or 4 of the 6 available 2$^3$S sublevels, $A_1$ to $A_6$ \cite{abboud_high_2004}.  Fig.~\ref{fig:levels} shows the 2$^3$S and 2$^3$P sublevel separations at 2\,T and the absorption peaks produced by selected transitions between them. 

Counterintuitively, the handedness of the pumping beam does not determine the polarity of the polarization created. While pumping f$_2^-$ drives -$m_F$ transitions from sublevels $A_5$ and $A_6$, negatively polarizing the gas, f$_4^-$ depopulates the $A_1$ through $A_4$ sublevels, resulting in positive net nuclear polarization, despite the negative circular polarization. Likewise, f$_2^+$ and f$_4^+$ generate positive and negative polarization, respectively \cite{nikiel_metastability_2007}. In this report, we present the magnitude of achieved polarizations.

The 2$^3$S sublevels chosen for pumping undergo strong populations changes and cannot be used to measure the polarization. Instead, a set of well-separated ``probe peaks'' are available at each handedness, appearing as weak but resolved pairs which serve as good indicators of sublevel populations \cite{suchanek_optical_2007}. Crucially, these are 2$^3$S sublevels which are not being pumped but are governed by metastability exchange---and thus spin temperature---so that they are representative of the polarization of the ground state population. Fig.~\ref{fig:levels} highlights the appropriate probe peaks for f$_2^-$ pumping. The f$_4$ lines require probe peaks of the same circular polarization, while the f$_2$ use the probe peaks of the opposite handedness.

\section{Results}
\begin{figure}
    \centering
    \includegraphics[width=\linewidth]{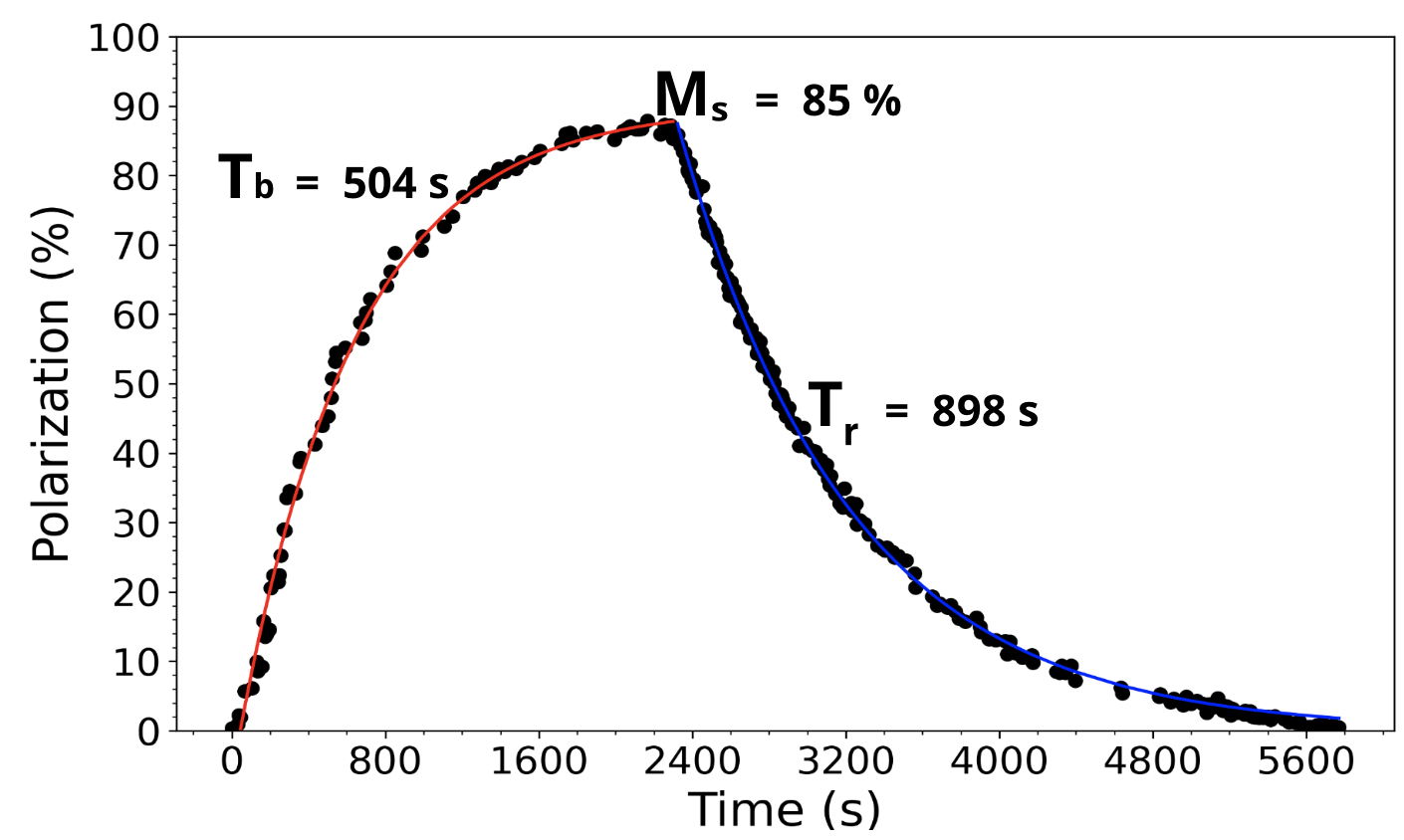}
    \caption{A representative pump and relaxation cycle at 5 T, featuring the f$_4^+$ pumping peak. The build-up time \( T_b \) and steady-state nuclear polarization \( M_{s} \) are extracted by fitting the optical pumping data from 0 s to 2300 s using the model described in \cite{gentile_spin-polarizing_1993}. The pump laser is deactivated at 2300 s. The relaxation time \( T_r \) is derived by fitting the exponential decay function to the data obtained subsequent to 2300 s.}
    \label{fig:generalpolvsrelax}
\end{figure}

Fig.~\ref{fig:generalpolvsrelax} illustrates a representative optical pumping and relaxation cycle for the polarization measurement of $^3$He in a 1.2\,mbar sealed cell at 5\,T, employing the f$_4^+$ pumping scheme. The pump laser is activated at 0\,s, initiating the build-up of nuclear polarization, which follows an exponential growth described by:
\begin{equation}
M(t) = M_{s} \left(1 - e^{-\frac{t}{T_b}}\right),
\label{buildupeqn} 
\end{equation}
where $M_{s}$ denotes the steady-state nuclear polarization and $T_b$ is the characteristic build-up time constant. At 2300\,s, the pump laser is deactivated, and the relaxation time $T_r$ is extracted by fitting the subsequent polarization decay data to an exponential function. The effective pumping rate $R$ is defined as $R = N M_{s}/T_b$,
where $N$ is the total number of atoms in the cell \cite{gentile_spin-polarizing_1993}. Systematic measurements of $M_s$, $T_b$, and $T_r$ were conducted at magnetic fields of 2, 3, and 5\,T to systematically evaluate the impact of discharge intensity, pump laser power, and optical pumping transition schemes on the performance of high-field metastability exchange optical pumping. 

\subsection{Pump Laser Power}
\label{laserpowerdependency}
\begin{figure}[h!]
    \centering
    \includegraphics[width=\linewidth]{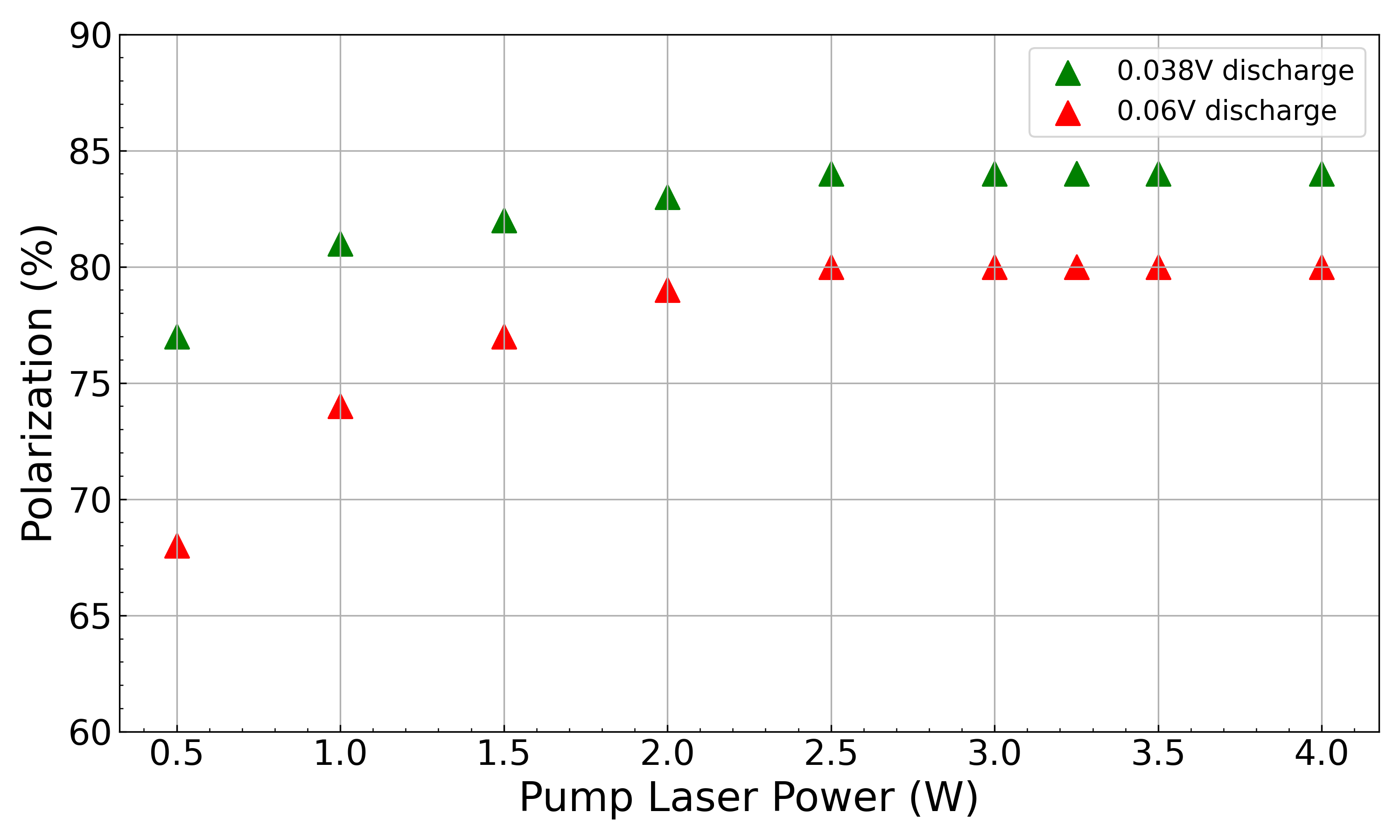}
    \caption{Maximum steady-state nuclear polarization measured at 5 T using the f$_4$$^+$ line for various pump laser power levels at two different discharge intensities.}
    \label{fig:pumplaser}
\end{figure}

Fig.~\ref{fig:pumplaser} illustrates the relationship between the steady-state nuclear polarization and the laser power for the 
f$_4^+$  transition scheme under a magnetic field of 5 T, evaluated for two distinct discharge amplitudes. While it is difficult to exactly express the quality of the discharge, we indicate these two intensity settings by their pre-amplified supply voltage as a rough guide. It's important to note that the pump laser output power is not the same as the power absorbed by the gas, or even the power incident on the portion of the gas volume lit with a discharge. This expanded beam was gaussian in profile, sending a significant portion of the laser power through the center of the cell where there was less discharge to absorb it.

The trend observed for two-sample discharge amplitudes indicates that polarization increases with pump laser power up to approximately 2.5 W, after which it appears to plateau, showing no significant dependence on discharge amplitude. The results reveal a maximum polarization of 84\% achieved using the  f$_4^+$
  peak with a 50 W RF amplifier, which is marginally lower (by 1\%) compared to the 85\% polarization attained with a 25 W RF amplifier, as depicted in Fig.~\ref{fig:generalpolvsrelax}. The high-power RF amplifier test was conducted to verify consistency with the results presented in Fig.~\ref{fig:generalpolvsrelax}, serving as a precursor to subsequent high-pressure tests. 
With the exception of magnetic fields below 1.5 T, the polarization demonstrated only a slight dependence on the magnetic field strength, consistently following the same trend for both discharge intensities of the sample.
\subsection{Discharge Intensity}
\begin{figure}[h!]
    \centering
    \includegraphics[width=\linewidth]{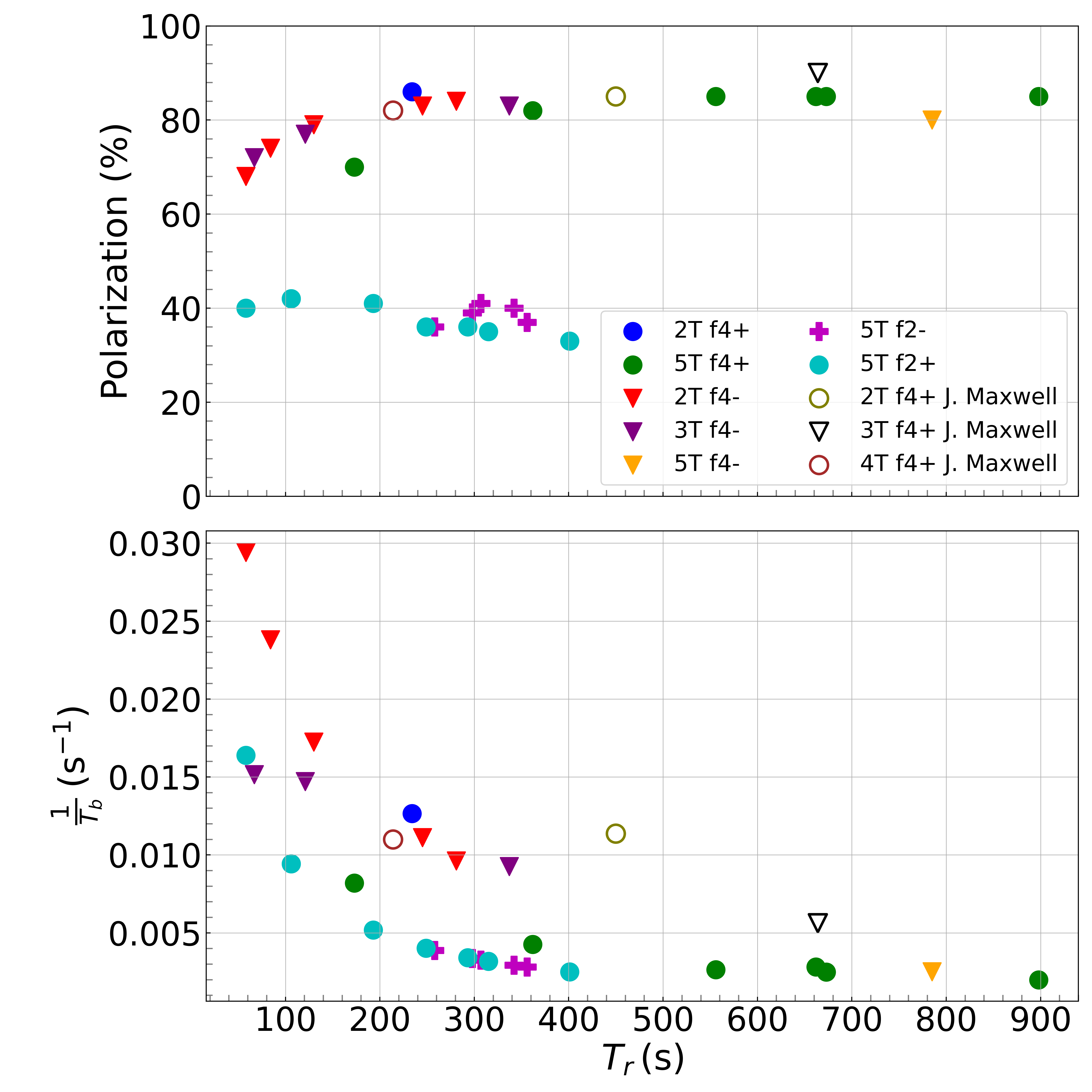}
    \caption{Steady-state nuclear polarization and pumping rate results versus relaxation time. Relaxation times (T$_r$) were varied by altering discharge intensity levels, within magnetic fields of 2 T, 3 T, and 5 T. The measurements utilized the f$_4^{\pm}$ and f$_2^{\pm}$ pumping schemes, with a constant pump laser output power of 3.25 W. Results on a
1 torr sealed cell produced at MIT Bates at 2 - 4 T are also included from \cite{maxwell_enhanced_2020}, for reference. The corresponding data points are distinguished by color and shape, as indicated in the legend.}
    \label{fig:polvsrel}
\end{figure}

The distribution and intensity of the discharge are significantly affected by the frequency and amplitude of the RF source, the electrode configuration, and the strength of the magnetic holding field. In this study, fine-tuning of the discharge intensity was achieved by systematically adjusting the voltage amplitude of the RF source. Optimal polarization performance was observed when the discharge was minimized to the lowest possible intensity while still fully illuminating the region intersected by the pumping laser. To achieve this condition, the experimental setup depicted in Fig.~\ref{fig:setup} was employed. The top panel of Fig.~\ref{fig:polvsrel} illustrates the steady-state nuclear polarizations attained at different relaxation times, T$_r$, which are predominantly determined by the RF discharge intensity. Notably, lower-intensity (dimmer) discharges result in substantially longer relaxation times.

 These optical pumping experiments utilized the f$_4^{\pm}$
  and f$_2^{\pm}$ pumping schemes, maintaining a constant pump laser output power of 3.25 W. The steady-state nuclear polarization for the f$_4$ pumping lines approached a maximum as the discharge dimmed and relaxation time increased. The longest relaxation time achieved, 898 s, at 5~T resulted in a steady-state polarization of 85\% using the f$_4^{+}$ pumping peak. The f$_2$ pumping line results were notably lower, around 40\%, and did not improve with increasing relaxation time. The change in the magnetic field in the tested range of 2 to 5\, T did not strongly affect the achieved polarization.
  As the maximum polarization increased, the polarization build-up rate (${T_b}^{-1}$) decreased accordingly, as shown in the bottom panel of Fig.~\ref{fig:polvsrel}. These findings align with results reported in references \cite{li_metastability_2023, maxwell_enhanced_2020}, both of which utilized a 1~torr $^3$He sealed cell. 

\subsection{Discussion}

Early results from the LKB which investigated the differences in the  f$_2$ and f$_4$ pumping lines \cite{abboud_high_2004, abboud_metastability_2005} stand somewhat at odds with the steady-state polarizations we are seeing. They found that while f$_4$ pumping gave much shorter build-up times, f$_2$ yielded higher steady-state polarizations due to higher photon efficiency---the number of nuclei polarized per photon absorbed. While f$_4$ rapidly empties four 2$^3$S sublevels, the cycle of absorption to collisional redistribution to spontaneous emission which results in polarization should be ultimately more efficient in the f$_2$ scheme where the angular momentum transfer is in the same direction as the desired polarization.

\begin{figure}[h!]
    \centering
    \includegraphics[width=1\linewidth]{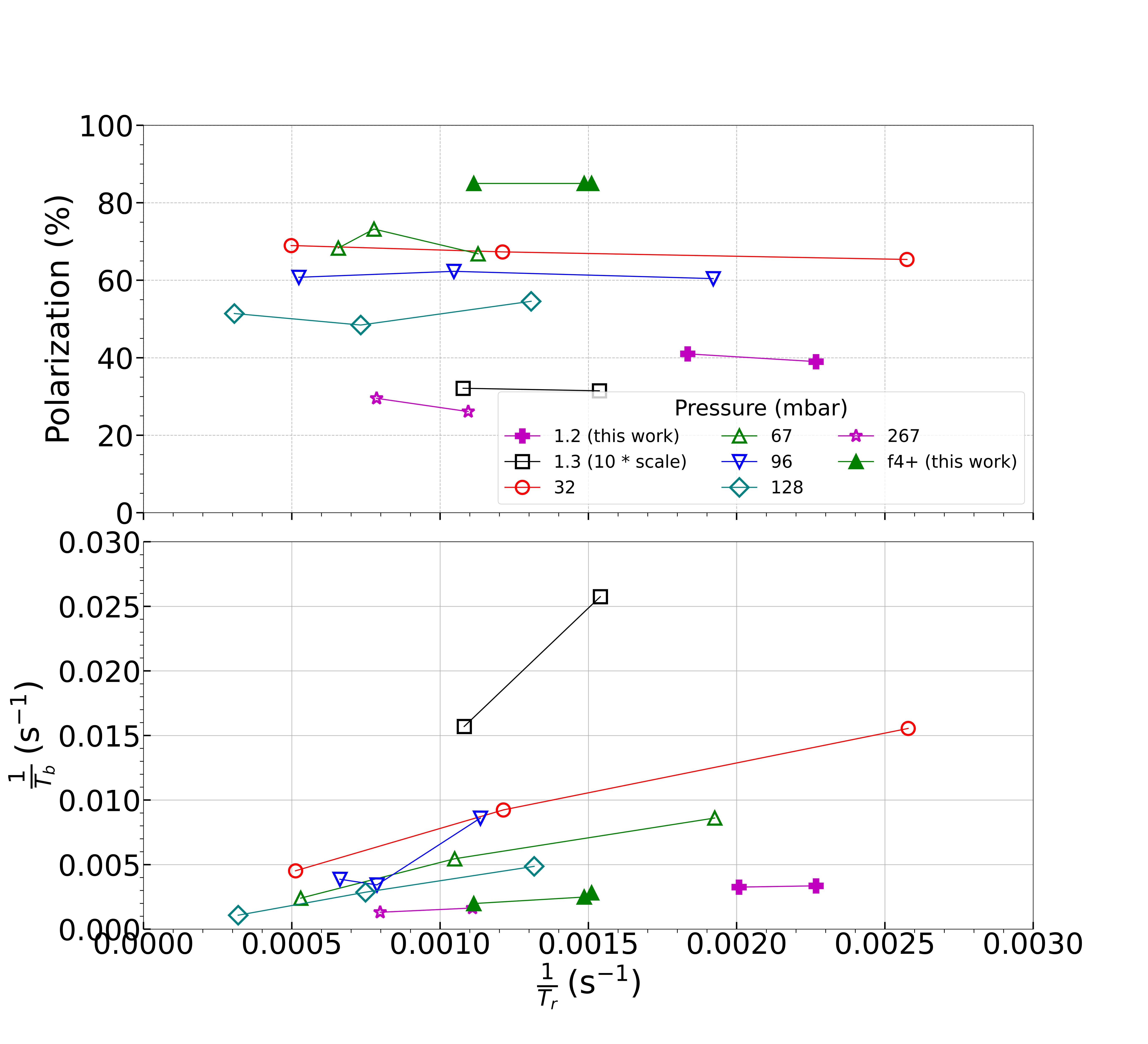}
  \caption{Polarization performance versus relaxation rate, predominantly reproduced from Nikiel et al. \cite{nikiel-osuchowska_metastability_2013}, with new points from our results for comparison. The filled green upward triangles and magenta pluses represent data from this work, showing the steady-state polarization as a function of the decay rate (top panel) and the build-up rate as a function of the relaxation rate (bottom panel) for the f$_2^-$ and f$_4^+$ optical pumping schemes. Measurements were performed at a laser power of 3.25 W, a magnetic field of 5 T, and a cell pressure of 1.2 mbar. All other data points correspond to results from Nikiel et al. \cite{nikiel-osuchowska_metastability_2013} under various plasma conditions, with a laser power of 0.5 W for the 4.7 T, f$_2^-$ OP line. Data for Nikiel’s 1.3 mbar results (black squares) are presented with a 10 times x-scale.}
    \label{fig:MGammaOPGammaD}
\end{figure}

Our results are directly comparable to the lowest pressure results from the LKB results reported in Nikiel-Osuchowska \textit{et al.} \cite{nikiel-osuchowska_metastability_2013}, and we see consistently improved performance at this pressure. The LKB effort was notably focused on the higher-pressure performance useful in medical imaging, and their 1.3 mbar cell seemed to be an outlier. Our results using the f$^-_2$ line at 1.2\,mbar are much more in-line with their higher pressure cell performance, in terms of maximum steady-state polarization achieved, and pumping and relaxation rates, as seen in Fig.~\ref{fig:MGammaOPGammaD}, where their 1.3\,mbar result is plotted on a ten times increased x-scale. While the 2013 LKB results focused on pumping the f$^-_2$ line, we saw much better results pumping the f$_4$ lines at these low pressures. 

Our achieved steady-state polarizations using the f$_4$ lines are exciting, though they make the much lower f$_2$ results more mysterious. The results from the LKB, as well as photon efficiency calculations, indicate that f$_2$ pumping should give higher polarizations, but we have consistently seen better results pumping f$_4$ in this and previous studies \cite{li_metastability_2023, maxwell_enhanced_2020}.

\section{Conclusions}
Our tests of metastability exchange optical pumping (MEOP) for the polarization of 1.3 mbar $^3$He in high magnetic fields from 2 to 5 \,T continue to support the feasibility of a proposed polarized $^3$He target for use in Hall B's CLAS12 detector system. Our results stray somewhat from established performance from the LKB, as our best results came from different pumping transitions, but the effectiveness of MEOP methods in high magnetic field environments remains compelling.  
Having established performance at these low pressures in a sealed cell, we will move to tests at higher pressures up to 100\,mbar, the goal pressure of our planned target system.

\section*{Acknowledgements}
The authors gratefully acknowledge the Jefferson Lab Target Group for their essential mechanical expertise and support. We also thank P.-J.~Nacher and G.~Tastevin of LKB, Paris, for helpful discussions and support. This work is supported by the U.S. Department of Energy, Office of Nuclear Physics, under grant numbers DE-FG02-94ER40818 awarded to the Massachusetts Institute of Technology and DE-AC05-06OR23177 awarded to Jefferson Lab.

\bibliographystyle{elsarticle-num}
\bibliography{references}

\end{document}